\documentclass[aip,apl,reprint,twocolumn]{revtex4-1}

\usepackage{graphicx}
\usepackage{dcolumn}
\usepackage{bm}
\usepackage{xfrac}
\usepackage{SIunits}
\usepackage{color}
\usepackage[usenames,dvipsnames]{xcolor}
\usepackage{hyperref}
\usepackage{footmisc}
\usepackage{amsmath}
\usepackage[normalem]{ulem}
\usepackage{natbib}
\usepackage{hhline}

\usepackage[utf8]{inputenc}
\usepackage[T1]{fontenc}
\usepackage{mathptmx}

\newcommand{\wc}{\omega_{\mathrm{res}}}
\newcommand{\geff}{g_{\mathrm{eff}}}
\newcommand{\keff}{\kappa_{\mathrm{eff}}}

\newcommand{\new}[1]{\textcolor{blue}{#1}}

\begin{document}

\title{Probing spin dynamics of ultra-thin van der Waals magnets via photon-magnon coupling}

\author{Christoph W. Zollitsch}
\email{c.zollitsch@ucl.ac.uk}
\affiliation{London Centre for Nanotechnology, University College London, 17-19 Gordon Street, London, WCH1 0AH, UK}
\author{Safe Khan}
\affiliation{London Centre for Nanotechnology, University College London, 17-19 Gordon Street, London, WCH1 0AH, UK}
\author{Vu Thanh Trung Nam}
\affiliation{Department of Physics, Faculty of Science, National University of Singapore, 2 Science Drive 3, Singapore 117542, Singapore}
\author{Ivan A. Verzhbitskiy}
\affiliation{Department of Physics, Faculty of Science, National University of Singapore, 2 Science Drive 3, Singapore 117542, Singapore}
\author{Dimitrios Sagkovits}
\affiliation{London Centre for Nanotechnology, University College London, 17-19 Gordon Street, London, WCH1 0AH, UK}
\affiliation{National Physical Laboratory, Hampton Road, Teddington TW11 0LW, UK}
\author{James O'Sullivan}
\affiliation{London Centre for Nanotechnology, University College London, 17-19 Gordon Street, London, WCH1 0AH, UK}
\author{Oscar W. Kennedy}
\affiliation{London Centre for Nanotechnology, University College London, 17-19 Gordon Street, London, WCH1 0AH, UK}
\author{Mara Strungaru}
\affiliation{Institute for Condensed Matter Physics and Complex Systems, School of Physics and Astronomy, The University of Edinburgh, Edinburgh EH9 3FD, UK}
\author{Elton J. G. Santos}
\affiliation{Institute for Condensed Matter Physics and Complex Systems, School of Physics and Astronomy, The University of Edinburgh, Edinburgh EH9 3FD, UK}
\affiliation{Higgs Centre for Theoretical Physics, The University of Edinburgh, Edinburgh EH9 3FD, UK}
\author{John J. L. Morton}
\affiliation{London Centre for Nanotechnology, University College London, 17-19 Gordon Street, London, WCH1 0AH, UK}
\affiliation{Department of Electronic \& Electrical Engineering, UCL, London WC1E 7JE, United Kingdom}
\author{Goki Eda}
\affiliation{Centre for Advanced 2D Materials, National University of Singapore, 6 Science Drive 2, Singapore 117546, Singapore}
\affiliation{Department of Physics, Faculty of Science, National University of Singapore, 2 Science Drive 3, Singapore 117542, Singapore}
\affiliation{Department of Chemistry, Faculty of Science, National University of Singapore, 3 Science Drive 3, Singapore 117543, Singapore}
\author{Hidekazu Kurebayashi}
\affiliation{London Centre for Nanotechnology, University College London, 17-19 Gordon Street, London, WCH1 0AH, UK}
\affiliation{Department of Electronic \& Electrical Engineering, UCL, London WC1E 7JE, United Kingdom}
\affiliation{WPI Advanced Institute for Materials Research, Tohoku University, 2-1-1, Katahira, Sendai, 980- 8577, Japan}

\date{\today}

\maketitle

\textbf{Layered van der Waals (vdW) magnets can maintain a magnetic order even down to the single-layer regime and hold promise for integrated spintronic devices. While the magnetic ground state of vdW magnets was extensively studied, key parameters of spin dynamics, like the Gilbert damping, crucial for designing ultra-fast spintronic devices, remains largely unexplored.
Despite recent studies by optical excitation and detection, achieving spin wave control with microwaves is highly desirable, as modern integrated information technologies predominantly are operated with these. The intrinsically small numbers of spins, however, poses a major challenge to this.
 Here, we present a hybrid approach to detect spin dynamics mediated by photon-magnon coupling between high-Q superconducting resonators and ultra-thin flakes of Cr$_2$Ge$_2$Te$_6$ (CGT) as thin as 11\,nm. We test and benchmark our technique with 23 individual CGT flakes and extract an upper limit for the Gilbert damping parameter. These results are crucial in designing on-chip integrated circuits using vdW magnets and offer prospects for probing  spin dynamics of monolayer vdW magnets.}

\section*{Introduction}
van der Waals (vdW) materials~\cite{Geim_Nature2013,Novoselov_Science2016,Genome22} consist of individual atomic layers bonded by vdW forces and can host different types of collective excitations such as plasmons, phonons and magnons. Strong coupling between these excitation modes and electromagnetic waves (i.e. photonic modes) creates confined light-matter hybrid modes, termed polaritons. Polaritons in vdW materials are an ideal model system to explore a variety of polaritonic states~\cite{Basov_Science2016,Low_Nature2017}, e.g. surface plasmon polaritons in graphene~\cite{Fei_Nature2012,Chen_Nature2012} and exciton polaritons in a monolayer MoS$_2$ embedded inside a dielectric microcavity~\cite{Liu_NPhoton2015}. These states can be further modified by electrostatic gating~\cite{Verzhbitskiy2020}, as well as by hetero-structuring with dissimilar vdW layers~\cite{Geim_Nature2013}. 

Numerous studies on magnon polaritons (MPs) \cite{Rameshti2022, Awschalom_IEEETQE2021} have been using macroscopic yttrium iron garnet (YIG) coupled to either three-dimensional cavities \cite{Zhang_PRL2014} or to on-chip resonators \cite{Huebl2013, Wang_PRL2019}, with potential applications in ultra-fast information processing, non-reciprocity or microwave to optical transduction. By reducing the number of excitations, MPs find application in the quantum regime e.g., magnon number counting via an electromagnetically coupled superconducting qubit \cite{Tabuchi_Science2015, LachanceQuirion2019} or as a building block for Bell state generation \cite{Yuan2022}.

The rapidly developing research around polaritons and specifically MPs has so far, been little studied in magnetic vdW materials due to the relatively recent discoveries of long-range magnetic order in vdW systems at the few monolayer regime \cite{Gong_Nature2017,Huang_Nature2017,Lee_NanoLett2016}, in addition to its technically challenging realization. Stable MP states are formed by strongly coupling the magnetic field oscillation of a resonant photon to the collective magnetization oscillation in a magnetic material. This strong coupling is achieved when the collective coupling rate $\geff$ is larger than the average of both system loss rates. In a simplified picture, $\geff$ scales linearly with the strength of the oscillating magnetic field of a resonator and the square root number of spins \cite{Huebl2013}. For studies involving bulk magnetic materials and low quality and large microwave resonators, strong coupling is achieved when $\geff/2\pi$ is in the MHz range, which is accomplished with relative ease due to the abundance of spins in bulk magnetic materials. A reduction of the bulk dimensions down from mm to $\mu$m and nm scales, the typical lateral dimensions and thickness of vdW material monolayers, results in a decrease of the coupling strength by at least 6 orders of magnitude. Commonly used microwave resonators are not able to produce strong enough oscillating magnetic fields to compensate for such a reduction in absolute number of spins. Only by advanced resonator design and engineering the regime of strongly coupled MPs in monolayer vdW magnetic materials can be accomplished, granting access to spin dynamic physics at a true 2d monolayer limit and research on MPs in nano-scale devices where the whole range of on-chip tuning and engineering tools, such as electric fields or device design, are available.

Magnons or magnon polaritons have been observed in magnetic vdW materials, but it had been restricted to either to the optical frequency range \cite{Zhang_NMater2020,Cenker_NPhys2021} or a large thickness limit \cite{Mandal2020, Zhang2021}, respectively. Here, we \new{present our attempt} of detecting spin dynamics in ultra-thin vdW magnetic materials and the creation of MPs by magnon-photon coupling in the microwave frequency range, using superconducting resonators optimized for increased magnon-photon coupling. By using microwave resonators with a small mode volume, we not only increased its oscillating magnetic field strength but also matched it more efficiently to the size of nanoscale vdW flakes. Our work presents a fundamental cornerstone for a general blueprint for designing and developing magnon-photon hybrids for any type of ultra-thin or monolayer vdW magnetic material, enabling research on on-chip microwave applications for (quantum) information processing.

%%%
\begin{figure}[!b]
 \includegraphics[]{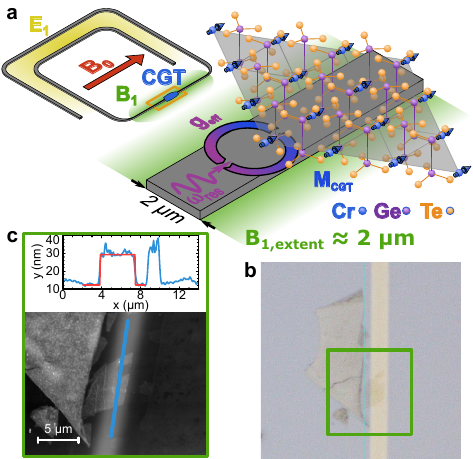}
 \caption{\textbf{Magnon-photon coupling between thin CGT and a superconducting resonator.} \textbf{a} Schematic of a resonator shows the design in detail, indicating the areas of high $E_1$-field (yellow) and $B_1$-field (green) intensities, as well as the orientation of the externally applied field $B_0$. Finally, a schematic zoom in of the section loaded with a CGT flake is shown. The collective coupling between a microwave photon and the magnetization of the CGT is illustrated, as well as the approximate extent of the microwave $B_1$-field. \textbf{b} Micrograph image of a CGT flake transferred onto the narrow section of a resonator. \textbf{c} AFM image of the CGT flake together with a height profile along the blue solid line in the AFM image. The red solid line is a fit to the flake thickness. The results of this resonator are presented in Fig.\,\ref{fig02}.
 \label{fig01}}
\end{figure}
%%%

%%%
\begin{figure*}[t]
 \includegraphics[]{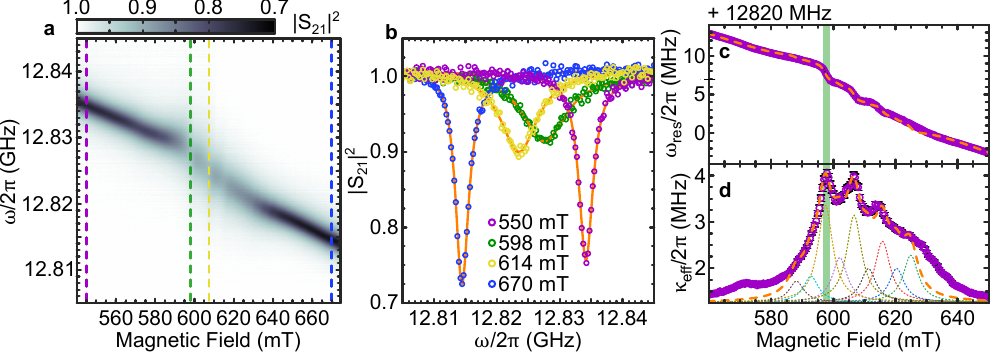}
 \caption{\textbf{Magnon-photon coupling observed in resonator microwave transmission.} \textbf{a} $\left|S_{21}\right|^2$ as a function static magnetic field $B_0$ and frequency, with the microwave transmission encoded in the color. The results are obtained from the resonator shown in Fig\,\ref{fig01}\,(b) and (c), featuring a loaded quality factor of $Q_\mathrm{L} = 4600$. \textbf{b} $\left|S_{21}\right|^2$ as a function of frequency at fixed magnetic fields, indicated in \textbf{a} by dashed vertical lines. \textbf{c} and \textbf{d} Resonance frequency $\wc$ and effective loss rate $\keff$ as a function of magnetic field. Note the multiple resonance peaks, indicating multiple CGT FMRs. The dashed orange lines are results from the semi-optimized fit. \textbf{d} exemplary includes the individual peaks of which the orange dashed lines consists. The green bar in \textbf{c} and \textbf{d} highlights the main mode.
 \label{fig02}}
\end{figure*}
%%%

\section*{Results}
In this article, we report on the observation of spin dynamics and the creation of MPs at the onset of the high cooperativity regime with the vdW ferromagnet CGT of $\nano\meter$ scale thickness, demonstrating a pathway towards stable magnon-photon polariton creation. We combine a precise transfer process of exfoliated CGT flakes and high sensitivity superconducting resonators, to access and study the dynamical response of coupled photon-magnon states in a small-volume ($\nano\meter$-thick and $\micro\meter$-sized) CGT flake (illustrated in Fig.\,\ref{fig01}\,(a)). High-quality-factor superconducting lumped element resonators are chosen to be the counterpart due to their extremely small mode volume ($\approx 6000\,\micro\meter^3$) and consequently strong oscillating magnetic fields ($B_1\,\approx 25\,\nano\tesla$, see SI for resonator quality-factors and $B_1$-field distributions), resulting in high spin sensitivities \cite{Eichler2017, Weichselbaumer2019}. At cryogenic temperatures, we perform low-power microwave spectroscopy on multiple resonator-vdW-flake hybrids, covering a frequency range from $12\,\giga\hertz$ to $18\,\giga\hertz$ for a variety of thickness. Samples consist of up to 12 resonators on a single chip, all capacitively coupled to a common microwave transmission line for read-out (see SI for details). Multiple peaks of spin-wave resonances are observed for each CGT flake measured. The spin-wave modes are closely spaced in frequency and show a large overlap. We employ a semi-optimized fitting model to produce a good estimate for the collective coupling strength and magnetic linewidth. By taking the resonance value of the most prominent peak of each spectrum, we find that all measured points can be fitted very well by a single curve calculated by the Kittel formula with bulk CGT parameters. Furthermore, we extracted the linewidth for the thinnest CGT flake investigated, $11\,\nano\meter$ or 15 monolayers (ML), the only device exhibiting well separated spin-wave modes. This allowed a fully quantitative analysis and we determined an upper limit of the Gilbert damping parameter of $0.02$. This value is comparable to the damping reported for 3d transition metal ferromagnets, suggesting that magnetic vdW flakes have the potential for the fabrication of functional spintronic devices. 

We investigate the dynamics of $\nano\meter$-thick CGT flakes, using superconducting lumped element resonators made of NbN (see methods for fabrication details and SI and Ref. [\onlinecite{Zollitsch2019}] for more performance details). The advantages of a lumped element design are the spatial separation of the oscillating magnetic field $B_1$ and electric field $E_1$ and the concentration of $B_1$ within a narrow wire section of the resonators, as indicated in Fig.\,\ref{fig01}\,(a). Additionally, the $B_1$ field distribution is homogeneous along the length of the narrow wire section (see finite element simulations in SI). This magnetic-field concentration is our primary reason to use this type of resonator in order to reduce the photon mode volume as well as achieve a considerable mode overlap between the resonator photon mode and CGT magnon mode, and consequently, a large coupling strength. We therefore transfer CGT flakes onto these narrow sections (Fig.\,\ref{fig01}\,(b)). Details of CGT flake transfers are described in the methods section. Optical imaging and atomic force microscopy (AFM) measurements are used to characterise the size and thickness of the CGT flakes (see Fig.\,\ref{fig01}\,(c)). Measured thicknesses range from $153 \pm 23\,\nano\meter$ down to $11 \pm 1.8\,\nano\meter$ (15 ML), enabling a thickness dependent study of CGT flakes and their coupling to the resonators. 

We measured the microwave transmission $\left| S_{21} \right|^2$ as a function of frequency and externally applied magnetic field $B_0$ for each resonator at a temperature of $1.8\,\kelvin$, using a microwave power of approximately $-80\,\deci\bel\meter$ at the resonator chip.
Figure\,\ref{fig02}\,(a) shows the resulting 2D plot of $\left|S_{21}\right|^2$ for a resonator loaded with a $17\,\nano\meter \pm 0.8\,\nano\meter$ thick CGT flake (see Fig.\,\ref{fig01}\,(b) and (c) for the respective micrograph and AFM images). A resonator peak can be clearly observed for each magnetic field, with its resonance frequency $\wc$ decreasing with increasing magnetic field. The reduction of the frequency is a result of a slow degradation of the superconductivity by $B_0$, which in general exhibits a parabolic dependence \cite{Healey2008}. For $580\,\milli\tesla \leq B_0 \leq 630\,\milli\tesla$ the resonator prominence is reduced, highlighted by $\left|S_{21}\right|^2$ as a function of frequency for four constant $B_0$ values in Fig.\,\ref{fig02}\,(b). Within this field range, the mode resonance has been modified due to its hybridization with the magnetic modes of the CGT flake. To further quantify the interaction, we fit each $\left|S_{21}\right|^2$ profile by a Fano resonance lineshape (solid orange lines in Fig.\,\ref{fig02}\,(b)) to account for an asymmetric resonance peak due to additional microwave interference in the circuitry \cite{Fano1961, Khalil2012},
%%%
\begin{equation}
    \left|S_{21}\right|^2 = S_0 + A \frac{\left( q\sfrac{\keff}{2} + \omega - \wc \right)^2}{(\sfrac{\keff}{2})^2 + \left( \omega - \wc \right)^2}.
\end{equation}
%%%
Here, $S_0$ is the microwave transmission baseline, $A$ the peak amplitude, $q$ describes the asymmetry of the lineshape and $\keff$ represents the effective loss rate of the hybrid system (see SI for resonator parameters before and after CGT transfer for all resonators). Figure\,\ref{fig02}\,(c) shows $\wc$ of the hybrid system as a function of $B_0$. $\wc$ experiences a dispersive shift when the photon mode and the magnon mode hybridize, indicating an onset of a strong interaction between the two individual systems \cite{Herskind2009, Bushev2011, Huebl2013, LachanceQuirion2019, Khan2021}. We observe multiple shifts in $\wc$, suggesting an interaction of several magnon modes with the resonator in our experiment. 

Signatures of the resonator--CGT-flake coupling are also characterised by $\keff$ of the hybrid system (Fig.\,\ref{fig02}\,(d)). $\keff$ is enhanced from the value of the resonator loss rate $\kappa_0$ due to an additional loss introduced by the magnon system characterized by the loss rate $\gamma$ \cite{Herskind2009, Huebl2013, Tabuchi2014}. Consistent with the $B_0$ dependence of $\wc$, $\keff$ shows a rich structure, having its main peak at $598\,\milli\tesla$, together with less prominent peaks distributed around it. 
Based on a formalism for coupled-harmonic-oscillator systems in the high cooperativity regime \cite{Herskind2009, Bushev2011, Khan2021}, we use the following to analyse our experimental results with multiple peaks: 
\begin{equation}
    \wc = \omega_{\mathrm{res,0}} + mB_0^2 + \sum_{k=-n}^{+n} \frac{g_{\mathrm{eff},k}^2 \Delta_k}{\Delta_k^2 + \gamma^2},
    \label{eq1:weff}
\end{equation}

\begin{equation}
    \keff = \kappa_0 + \sum_{k=-n}^{+n} \frac{g_{\mathrm{eff},k}^2 \gamma}{\Delta_k^2 + \gamma^2}.
    \label{eq2:keff}
\end{equation}
with the detuning factor for each resonance as $\Delta_k = \frac{g_{\mathrm{CGT}}\mu_{\mathrm{B}}}{\hbar} \left( B_0 - B_{\mathrm{FMR}, k} \right)$.
%%%
\begin{figure}[b]
 \includegraphics[]{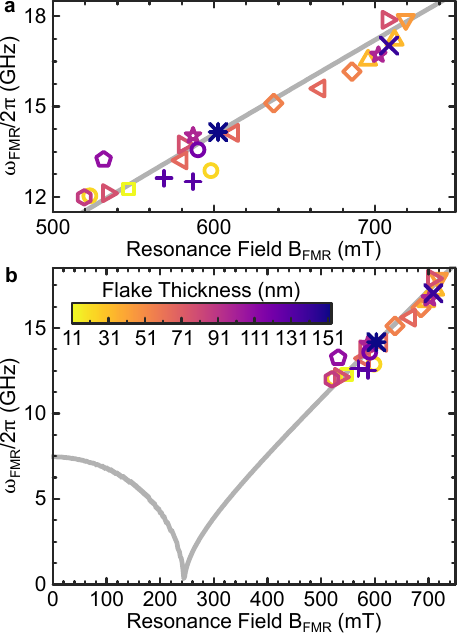}
 \caption{\textbf{Summary of CGT-FMR conditions.} \textbf{a} Extracted CGT resonance fields and frequencies from the set of resonators loaded with CGT flakes of different thickness. Resonance values are taken from the most prominent peaks in $\keff$. The solid curve is calculated using the Kittel formalism presented in \cite{Khan2019}, using same parameters, with $g_{\mathrm{CGT}} = 2.18$, $\mu_0M_{\mathrm{s}} = 211.4\,\milli\tesla$ and $K_{\mathrm{u}} = 3.84 \times 10^4\,\sfrac{\joule}{\meter^3}$. \textbf{b} Wider magnetic field range of \textbf{a} where the CGT flake thickness for the different symbols is indicated by the color gradient given in \textbf{a}.
 \label{fig03}}
\end{figure}
%%%
Here, $\omega_{\mathrm{res,0}}$ is the resonator resonance frequency at $B_0 = 0\,\tesla$ and $m$ represents the curvature of the resonance frequency decrease due to the applied magnetic field. $B_{\mathrm{FMR}, k}$ is the CGT FMR field, $g_{\mathrm{CGT}}$ the g-factor of CGT and $g_{\mathrm{eff},k}$ gives the collective coupling strength between photon and magnon mode. The summation is over all resonance modes $k$ present on the low or high field (frequency) side of the main resonance mode, where $n$ gives the number of modes on one side. For simplicity, we assume a symmetric distribution of modes about the main mode. The large number of multiple modes and their strong overlap prevent a reliable application of a fully optimized fit to the data, due to the large number of free parameters required. In an effort to gain a good estimate of the model parameters we apply the model functions Eq.\,(\ref{eq1:weff}) and (\ref{eq2:keff}) in a two-step semi-optimized fashion (see SI for details).
With this approach, we arrive at a model in good agreement with $\wc$ and $\keff$ (see orange dashed lines in Fig.\,\ref{fig02}\,(c), (d), exemplary showing the individual peaks of the orange dashed line in Fig.\,\ref{fig02}\,(d) and the SI for additional results and data). We can reproduce the data using $\gamma/2\pi = 94.03 \pm 5.95\,\mega\hertz$ and a collective coupling strength of the main mode of $13.25 \pm 1\,\mega\hertz$. Together with $\kappa_0/2\pi = 1.4 \pm 0.02\,\mega\hertz$ the system resides at the onset of the high cooperativity regime, classified by the cooperativity $C = \sfrac{\geff^2}{\kappa_0\gamma} = 1.3 > 1$ \cite{Herskind2009, Zhang_PRL2014}. In this regime, magnon polaritons are created and coherently exchange excitations between magnons and resonator photons on a rate given by $\geff$. The created MPs are, however, short lived and the excitations predominately dissipate in the magnonic system, as $\geff \ll \gamma$. 

Our analysis suggests that the separation of the different FMR modes is of the same order of magnitude as the loss rate (see SI for additional data). We consider that these are from standing spin wave resonances, commonly observed for thin magnetic films~\cite{Serga2010} and with one reported observation in bulk of the vdW material CrI$_3$ \cite{Kapoor2021}. In thin-film magnets under a static magnetic field applied in-plane, the magnetic-dipole interaction generates two prominent spin wave branches for an in-plane momentum, the backward volume spin wave (BVMSW) and magnetostatic surface spin wave (MSSW) modes \cite{Kalinikos_JPC1986, Bhaskar2020}. These spin wave modes have different dispersion relations, having higher (MSSWs) and lower (BVMSWs) resonance frequencies with respect to that of the uniform FMR mode. We calculate the distance of these standing spin-wave modes based on magnetic parameters of bulk CGT as well as the lateral dimensions of the flakes (see SI for more details). We can find spin waves having a frequency separation within $100\,\mega\hertz$ and $200\,\mega\hertz$ ($3.3\,\milli\tesla$ to $6.6\,\milli\tesla$ in magnetic field units), which are consistent with our experimental observation in terms of its mode separation. However, the irregular shape of the CGT flakes renders exact calculations of spin wave mode frequencies very challenging.
We also considered a possibility that each layer of CGT might have different magnetic parameters (e.g. chemical inhomogeneity), and thus producing different individual resonance modes.  Our numerical simulations based on atomistic spin dynamics\cite{wahab2021quantum,kartsev2020biquadratic} rule out this possibility, as resonance modes from individual layers average to a single mode as soon as a fraction of $10\%$ of inter-layer exchange coupling is introduced (see SI for more details). Therefore, we speculate that the multiple mode nature we observe in our experiments is likely originating from intrinsic properties of the CGT flakes. 

Figure\,\ref{fig03} shows the extracted $\omega_{\mathrm{FRM}}$ as a function of $B_{\mathrm{FMR}}$ for each resonator--CGT-flake hybrid. The experimental values are in excellent agreement with a curve calculated by the Kittel equation with magnetic parameters for bulk CGT~\cite{Khan2019}, from which the data exhibits a standard deviation of less than $5\,\%$. This agreement, achieved by independent characterizations of 23 CGT flakes measured by superconducting resonators, is experimental evidence that the magnetic parameters that determine the dispersion of $\omega_{\mathrm{FRM}}\left(B_{\mathrm{FMR}}\right)$, i.e. the CGT g-factor $g_{\mathrm{CGT}}$, saturation magnetization $M_{\mathrm{s}}$ and uniaxial anisotropy $K_{\mathrm{u}}$, exhibit little thickness dependence in exfoliated CGT flakes, and are not disturbed by the transfer onto the resonator structure. We note, that this demonstrates that vdW magnetic materials are particularly attractive for device applications, as they are less prone for contamination from exfoliation. 

\begin{figure}[t]
 \includegraphics[]{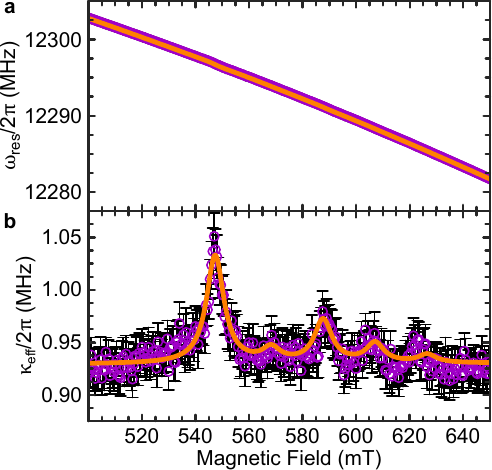}
\caption{\textbf{Magnon-photon coupling for the thinnest CGT flake.} \textbf{a} Resonance frequency $\wc$ and \textbf{b} effective loss rate $\keff$ as a function of magnetic field of a resonator loaded with the thinnest CGT, consisting of 15 ML. The resonator's loaded quality factor is 6938. The solid orange lines are results a fit to Eq.(\ref{eq1:weff}) and (\ref{eq2:keff}), respectively. The errorbars in \textbf{b} represent the standard deviation from the Fano resonance lineshape fit to the resonator transmission.
 \label{fig04}}
\end{figure}

Finally, we present our analysis of $\keff$ for a resonator with a $11 \pm 1.8\,\nano\meter$ CGT flake in Fig.\ref{fig04}. With the thickness of a single layer of CGT being $0.7\,\nano\meter$ \cite{Gong_Nature2017}, this flake consists of 15 monolayers and is the thinnest in our series. Figure \ref{fig04}\,(a) and (b) show $\wc$ and $\keff$ as a function of $B_0$, respectively. While the response of the CGT flake shows a prominent signature in $\keff$, the CGT FMR is considerably more subtle in $\wc$. This highlights the excellent sensitivity of the high-Q superconducting resonators in our study. $\keff$ features five well-separated peaks with the main peak at $B_0 = 547\,\milli\tesla$, which enables us to perform a single-peak fully optimized analysis for each, in contrast to our multi-step analysis for the remainder of the devices. We assume the additional peaks are BVMSW modes, as discussed in the previous section. However, the splitting is about four times larger than compared to all other investigated devices, which would result in a significantly shorter wavelength. Thickness steps can lead to a wavelength down-conversion \cite{Stigloher2018}, however, due to the irregular shape and $B_1$ inhomogeneities it is difficult to exactly calculate the spin wave frequencies (see SI for further details).
From the main peak profile, we extract $\geff/2\pi = 3.61 \pm 0.09\,\mega\hertz$, $\gamma/2\pi = 126.26 \pm 8.5\,\mega\hertz$ and $\kappa_0/2\pi = 0.92 \pm 0.05\,\mega\hertz$. We compare the experimental value of $\geff$ with a numerically calculated $g_\mathrm{eff,simu}$, using the dimensions of the CGT flake determined by AFM measurements (see SI for details). The calculation yields $g_\mathrm{eff,simu}/2\pi = 8.94\,\mega\hertz$, lying within the same order of magnitude. The overestimation is likely due to in-perfect experimental conditions, like non-optimal placement of the flake, uncertainties in the thickness and dimension determination as well as excluding the additional modes in the calculation (see SI). With $\gamma \gg \geff$ and $C = 0.11$, the hybrid system is in the weak coupling regime \cite{Zhang_PRL2014}, but due to the highly sensitive resonator with its small $\kappa_0$ the response from the magnon system can still be detected.
With the extracted $\gamma/2\pi$ we can give an upper limit of the Gilbert damping in CGT, by calculating $\alpha_\mathrm{upper} = \gamma/\omega_\mathrm{FMR}$. We find $\alpha_\mathrm{upper}$ as $0.021 \pm 0.002$, which is comparable to other transition metal magnetic materials \cite{Mankovsky_PRB2013}, and is in very good agreement with a previously reported effective Gilbert damping parameter determined by laser induced magnetization dynamics \cite{Zhang2020}. Here, we emphasise that the actual Gilbert damping value is lower due to a finite, extrinsic inhomogeneous broadening contribution. 

We further use these results to benchmark the sensitivity of our measurement techniques. The detection limit is given by comparing the main peak height characterised by $\geff^2/\gamma$ and the median noise amplitude which is $18\,\kilo\hertz$ in Fig.\,\ref{fig04}\,(b) where  $\geff^2/2\pi\gamma$ = 103 kHz. By assuming the same lateral dimensions and scale the thickness down to a single monolayer, while keeping $\gamma$ constant, we calculate the expected signal reduction numerically by $g_\mathrm{eff,simu,1ML}/g_\mathrm{eff,simu,15ML}$ to 0.26.
We obtain $(0.26\geff)^2/2\pi\gamma = 7\,\kilo\hertz$ for the monolayer limit. Although this suggests the noise amplitude is greater than the expected peak amplitude, we can overturn this condition by improving the coupling strength by optimising the resonator design, enhancing the exfoliation and flake transfer as well as by reducing the noise level by averaging a number of multiple scans. 
Superconducting resonators with mode volumes of about $10\,\micro\meter^3$ have been realised~\cite{Probst2017}, a reduction of 2 orders of magnitude compared to our current design. This would translate to an order of magnitude improvement in $\geff$. Furthermore, this flake covers about $4\,\%$ of the resonator. By assuming maximised coverage a 5 times enhancement of $\geff$ can be achieved. Both approaches would make the detection of monolayer flakes possible.

%%%
%%%

In summary, we provide the first demonstration of photon-magnon coupling between a superconducting resonator and $\nano\meter$-thick vdW flakes of CGT, using a total of 23 devices with different CGT flakes of thickness from $153\,\nano\meter$ down to $11\,\nano\meter$. By employing a coupled-harmonic-oscillator model, we extract the coupling strength, magnetic resonance field and relaxation rates for both photon and magnon modes in our devices. From our semi-broadband experiments, we find that the magnetic properties of exfoliated CGT flakes are robust against the transfer process, with a standard deviation of less than $5\,\%$ to expected resonance values from bulk parameters. Notably, this suggests that vdW magnetic materials can be pre-screened at bulk to identify the most promising material for few layer device fabrication. The upper limit of the Gilbert damping in the 15 ML thick CGT flake is determined to be $0.021$, which is comparable to commonly used ferromagnetic thin-films such as NiFe and CoFeB and thus making CGT attractive for similar device applications. We highlight that the damping parameter is key in precessional magnetisation switching \cite{Rowlands2019, Meo2022}, auto-oscillations by dc currents \cite{Wagner2018, Haidar2019}, and comprehensive spin-orbit transport in vdW magnetic systems~\cite{Kurebayashi_NatRevPhys2022}. The presented techniques are readily transferable to other vdW magnetic systems to study spin dynamics in atomically-thin crystalline materials. While creating stable magnon polaritons is still an open challenge due to the large loss rate $\gamma$ of the CGT magnon system, this work offers an important approach towards its achievement. There are still potential improvements to the measurement sensitivity such as resonator mode volume reduction by introducing $\nano\meter$ scale constrictions\cite{McKenzie2019, Gimeno2020} and use of exfoliation/transfer techniques to produce larger flakes to enhance the mode overlap (hence coupling strength)\cite{Huang2020, Zhou2022}. With concerted efforts, the formation of magnon polaritons in few layers vdW materials will become feasible.

\section*{Methods}
\textbf{Superconducting Resonators:} The resonators were fabricated by direct laser writing and a metal lift-off process. The individual $5\,\milli\meter \times 5\,\milli\meter$ chips are scribed from an intrinsic, high resistivity ($\rho > 5000\,\ohm\centi\meter$) n-type silicon wafer of $250\,\micro\meter$ thickness. For a well defined lift-off, we use a double photoresist layer of LOR and SR1805. The resonator structures are transferred into the resist by a Heidelberg Direct Writer system. After development, $\sim 50\,\nano\meter$ NbN are deposited by magnetron sputtering in a SVS6000 chamber, at a base pressure of $7 \times 10^{-7}\,$mbar, using a sputter power of $200\,\watt$ in an 50:50 Ar/N atmosphere held at $5 \times 10^{-3}\,$mbar, with both gas flows set to 50 SCCM \cite{Zollitsch2019}. Finally, the lift-off is done in a 1165 solvent to release the resonator structures. 

\textbf{CGT Crystal Growth:} CGT crystals used in this study were grown via chemical vapour transport. To this end, high-purity elemental precursors of Cr (chips, $\geq\,99.995\%$), Ge (powder, $\geq\,99.999\%$), and Te (shots, $99.999\%$) were mixed in the molar weight ratio Cr:Ge:Te = 10:13.5:76.5, loaded into a thick-wall quartz ampule and sealed under the vacuum of $\sim 10^{-5}\,$mbar. Then, the ampule was loaded into a two-zone furnace, heated up and kept at $950\,\degreecelsius$ for 1 week to homogenize the precursors. To ensure high-quality growth, the ampule was slowly cooled ($0.4\,\degreecelsius/\hour$) maintaining a small temperature gradient between the opposite ends of the ampule. Once the ampule reached $500\,\degreecelsius$, the furnace was turned off allowing the ampule to cool down to room temperature naturally. The large ($\sim 1\,\centi\meter$) single-crystalline flakes were extracted from the excess tellurium and stored in the inert environment.

\textbf{CGT Flake Transfer:} Devices for this study were made via transfer of single-crystalline thin flakes on top of the superconducting resonators. The flakes were first exfoliated from bulk crystals on the clean surface of a home-cured PDMS (polydimethylsiloxane, Sylgard 184) substrate.  The thickness of the CGT flakes on PDMS was estimated through the contrast variation with transmission optical microscopy. Then, the selected flake was transferred to a resonator. The transfer was performed in air at room temperature. To minimize the air exposure, the entire process of exfoliation, inspection and transfer was reduced to 10-15 min per resonator. For the flakes thicker than $50\,\nano\meter$, the strong optical absorption of CGT prevented the accurate thickness estimation with optical contrast. For those flakes, the thickness was estimated via a quick AFM scan performed on the PDMS substrate before the transfer step. Ready devices were stored in inert conditions.

\section*{Data Availability}

The data that support the findings of this study are available within the paper, Supplemental Material and from the corresponding authors upon reasonable request.  

\section*{References}
%\bibliography{CGT_Linewidth}

\section*{Acknowledgment}
This study is supported by EPSRC on EP/T006749/1 and EP/V035630/1. 
G.E. acknowledges support from the Ministry of Education (MOE), Singapore, under AcRF Tier 3 (MOE2018-T3-1-005) and the Singapore National Research Foundation for funding the research under medium-sized centre program.
E.J.G.S. acknowledges computational resources through CIRRUS Tier-2 HPC 
Service (ec131 Cirrus Project) at EPCC (http://www.cirrus.ac.uk) funded 
by the University of Edinburgh and EPSRC (EP/P020267/1); 
ARCHER UK National Supercomputing Service (http://www.archer.ac.uk) {\it via} 
Project d429. E.J.G.S acknowledges the Spanish Ministry of 
Science's grant program ``Europa-Excelencia'' under 
grant number EUR2020-112238, the EPSRC Early Career 
Fellowship (EP/T021578/1), and the University of 
Edinburgh for funding support.
D.S. acknowledges EPSRC funding through the Centre for Doctoral Training in
Advanced Characterisation of Materials (EP/L015277/1) and European Union’s Horizon 2020 Research and Innovation program under grant agreement GrapheneCore3, number 881603 and the Department for Business, Energy and Industrial Strategy through the NPL Quantum Program.

\section*{Author Contribution}
C.W.Z, S.K. and H.K. conceived the experimental project. Resonator design and optimization was done by J.O’S., O.W.K, C.W.Z and supervised by J.J.L.M. Resonator fabrication and characterization was done by C.W.Z. CGT crystals were grown by I.A.V. and exfoliated and transferred by I.A.V. and N.V.T.T. and supervised by E.G.. D.S. measured AFM on the CGT flakes on the resonators. C.W.Z. performed the experiments and the data analysis with input from S.K. and H.K. Atomistic spin dynamics simulations were carried out by M.S. supervised by E.J.G.S.. C.W.Z., M.S., I.A.V. and H.K. wrote the manuscript with input from all authors.

\section*{Competing interests}

The Authors declare no conflict of interests.

\pagebreak
\widetext
\begin{center}
\textbf{\large Supplemental Material - Probing spin dynamics of ultra-thin van der Waals magnets via photon-magnon coupling}
\end{center}

\setcounter{equation}{0}
\setcounter{figure}{0}
\setcounter{table}{0}
\setcounter{page}{1}
\makeatletter

\renewcommand{\thefigure}{S\arabic{figure}}
\renewcommand{\theequation}{S\arabic{equation}}
\renewcommand{\bibnumfmt}[1]{[S#1]}
\renewcommand{\citenumfont}[1]{S#1}

\newcommand{\lL}{\lambda_\mathrm{L}}

\author{Christoph W. Zollitsch}
\email{c.zollitsch@ucl.ac.uk}
\affiliation{London Centre for Nanotechnology, University College London, 17-19 Gordon Street, London, WCH1 0AH, UK}
\author{Safe Khan}
\affiliation{London Centre for Nanotechnology, University College London, 17-19 Gordon Street, London, WCH1 0AH, UK}
\author{Vu Thanh Trung Nam}
\affiliation{Department of Physics, Faculty of Science, National University of Singapore, 2 Science Drive 3, Singapore 117542, Singapore}
\author{Ivan A. Verzhbitskiy}
\affiliation{Department of Physics, Faculty of Science, National University of Singapore, 2 Science Drive 3, Singapore 117542, Singapore}
\author{Dimitrios Sagkovits}
\affiliation{London Centre for Nanotechnology, University College London, 17-19 Gordon Street, London, WCH1 0AH, UK}
\affiliation{National Physical Laboratory, Hampton Road, Teddington TW11 0LW, UK}
\author{James O'Sullivan}
\affiliation{London Centre for Nanotechnology, University College London, 17-19 Gordon Street, London, WCH1 0AH, UK}
\author{Oscar W. Kennedy}
\affiliation{London Centre for Nanotechnology, University College London, 17-19 Gordon Street, London, WCH1 0AH, UK}
\author{Mara Strungaru}
\affiliation{Institute for Condensed Matter Physics and Complex Systems, School of Physics and Astronomy, The University of Edinburgh, Edinburgh EH9 3FD, UK}
\author{Elton J. G. Santos}
\affiliation{Institute for Condensed Matter Physics and Complex Systems, School of Physics and Astronomy, The University of Edinburgh, Edinburgh EH9 3FD, UK}
\affiliation{Higgs Centre for Theoretical Physics, The University of Edinburgh, Edinburgh EH9 3FD, UK}
\author{John J. L. Morton}
\affiliation{London Centre for Nanotechnology, University College London, 17-19 Gordon Street, London, WCH1 0AH, UK}
\affiliation{Department of Electronic \& Electrical Engineering, UCL, London WC1E 7JE, United Kingdom}
\author{Goki Eda}
\affiliation{Centre for Advanced 2D Materials, National University of Singapore, 6 Science Drive 2, Singapore 117546, Singapore}
\affiliation{Department of Physics, Faculty of Science, National University of Singapore, 2 Science Drive 3, Singapore 117542, Singapore}
\affiliation{Department of Chemistry, Faculty of Science, National University of Singapore, 3 Science Drive 3, Singapore 117543, Singapore}
\author{Hidekazu Kurebayashi}
\affiliation{London Centre for Nanotechnology, University College London, 17-19 Gordon Street, London, WCH1 0AH, UK}
\affiliation{Department of Electronic \& Electrical Engineering, UCL, London WC1E 7JE, United Kingdom}
\affiliation{WPI Advanced Institute for Materials Research, Tohoku University, 2-1-1, Katahira, Sendai, 980- 8577, Japan}

\date{\today}

\maketitle

\section{Microwave Setup and Measurement}

%%%
\begin{figure}[!h]
 \includegraphics[]{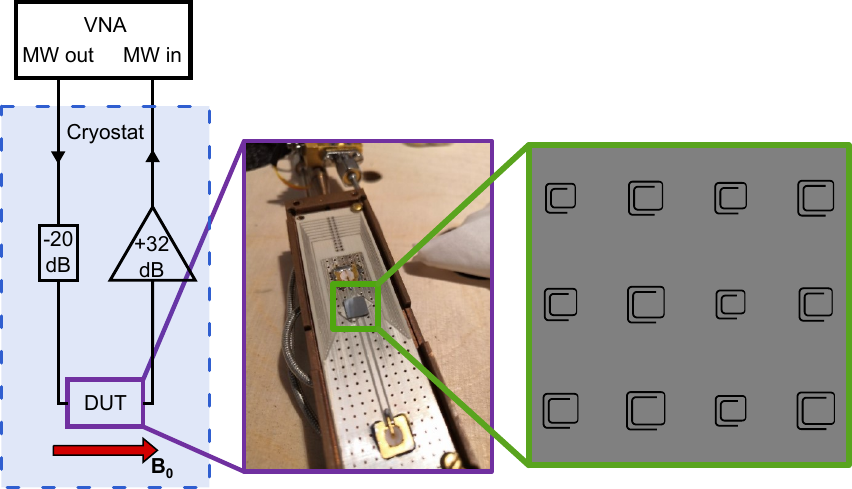}
 \caption{\textbf{Microwave delivery and detection setup.} Schematic of the microwave delivery and detection circuit. The image shows the coplanar waveguide transmission line. A resonator chip is placed on top of the transmission line for read out. On the right, a schematic layout of the resonators on a single chip is shown. 
 \label{S1}}
\end{figure}
%%%
Figure\,\ref{S1} shows a schematic of the used microwave measurement setup. We are using a Keysight E5071C vector network analyzer (VNA) to deliver and detect microwaves. The VNA is connected to a low temperature probe, fitted into a closed cycle helium cryostat and cooled to a base temperature of about $1.8\,\kelvin$. The microwave signal is transmitted into the cryostat and is attenuated by $-20\,\deci\bel$. The attenuator is positioned just before the sample box and provides a thermal anchoring for the center conductor of the coaxial cable to minimize the thermal load onto the sample. The output line is equipped with a Low Noise Factory LNC6\_20C cryogenic amplifier, operating between $6 - 20\,\giga\hertz$ with an average amplification of $+32\,\deci\bel$. The transmitted and amplified signal is finally detected by the VNA. Figure\,\ref{S1} also shows an image of the coplanar waveguide transmission line PCB, loaded with a resonator ship, of which a schematic shows the resonator layout on a single chip. The resonators on the chip are capacitively coupled to the transmission line PCB. Upon resonance the transmission through the PCB is reduced, indicating the resonator resonance. The cryostat is equipped with a mechanical rotation stage and prior to the measurements the superconducting resonators are carefully aligned to the externally applied static magnetic field $B_0$, such that the field is in the plane of the superconductor and along the narrow section of the resonators.

\begin{figure}[!h]
 \includegraphics[]{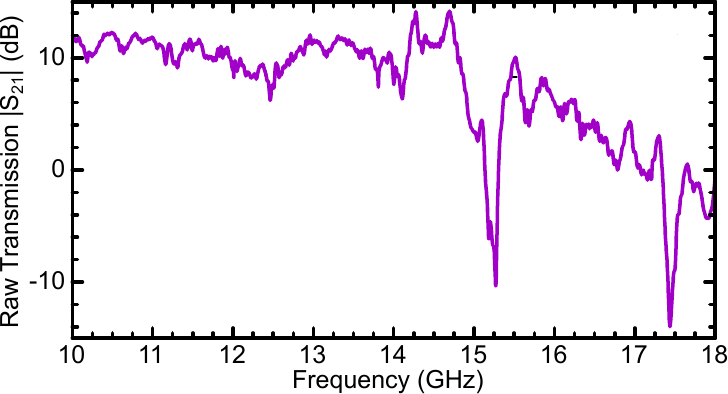}
 \caption{\textbf{Raw broadband microwave transmission signal.} Logarithmic microwave transmission $\left| S_{21} \right|$ as a function of frequency between $10\,\giga\hertz$ and $18\,\giga\hertz$ at a temperature of $1.8\,\kelvin$. 
 \label{S1.2}}
\end{figure}

Figure \ref{S1.2} shows the raw uncalibrated microwave transmission, ranging from $10\,\giga\hertz$ to $18\,\giga\hertz$. The transmission is dominated by imperfections in our microwave circuitry, masking the small signals from the superconducting resonators. Thus, we performed a simple thru calibration of the microwave transmission to remove contributions from the setup, prior each magnetic field dependent measurement. Here, we exploit the magnetic field tunability of our superconducting resonators. Before calibration, we set the frequency range of the measurement. We change the applied magnetic field such that the resonator's resonance frequency is tuned out of the set frequency range. With a frequency window just showing the transmission of the setup we perform the thru calibration. After calibration we set the magnetic field back to its starting value, resulting in a background corrected spectrum with just the resonator feature on it.

%%%
%%%

\section{Resonator Characterization}

In this study, we fabricate twelve superconducting lumped element resonators on each of three resonator chips were fabricated using the same design (see schematic Fig.\,1\,(a) in the main text). %Due to a finite fabrication yield not all of the 12 resonators were working on the chips. 
Prior to transfer of the CGT flakes, we characterized the resonators at a temperature of $1.8\,\kelvin$ and zero applied magnetic field, using microwave powers of about $-80\,\deci\bel\milli$ at the resonators, which is well below the bifurcation limit starting above $-60\,\deci\bel\milli$. Due to finite fabrication tolerances the resonator parameters have some variation, while some didn't work at all. However, the targeted resonance frequencies are well reproducible and very similar for the 3 different chips. We compare the resonator parameters before and after transfer of the CGT flakes and collate the parameters in Tab.\,\ref{tab1}. Note, the resonator parameters with the CGT flakes on were obtained with a static magnetic field applied in the plane of the superconductor, but far detuned from the CGT FMR. In addition, we add the respective thickness of the flake on each resonator, acquired from AFM measurements. Here, we give the values of the thickest region of a given flake on a resonator, as the thickest region will dominate the FMR signal. Due to the arbitrary shape of exfoliated flakes, some exhibit regions of different thickness, as seen e.g. in Fig.\,\ref{S4}\,(h) and (i).
\begin{table}[!ht]
    \centering
    \caption{Resonator Parameters}
    \begin{tabular}{|c|c|c|c|c|c|}
    \hline
        Chip Number & $\omega_{\mathrm{res, before}}$ ($\mega\hertz$) & $Q_{\mathrm{L, before}}$ & $\omega_{\mathrm{res, after}}$ ($\mega\hertz$) & $Q_{\mathrm{L, after}}$ & CGT Thickness ($\nano\meter$) \\
    \hline
        1 & 12165 & 1978 & 12063 & 5733 & 16.2 $\pm$ 1.3 \\
    \hline
        1 & 13303 & 7357 & 13177 & 4950 & - \\
    \hline
        1 & 13968 & 5575 & 13860 & 4679 & 49.4 $\pm$ 3.5 \\
    \hline
        1 & 14184 & 6492 & 14048 & 5627 & 153.1 $\pm$ 23.3 \\
    \hline
        1 & 16648 & 6606 & 16470 & 5021 & 23.5 $\pm$ 2.5 \\
    \hline
        1 & 17431 & 3215 & 17237 & 6826 & 23.8 $\pm$ 6.4 \\
    \hline
        1 & 17959 & 7595 & 17790 & 3963 & 26.2 $\pm$ 4.1 \\
    \hhline{|=|=|=|=|=|=|}
        2 & 12285 & 360 & 12153 & 7135 & 49.1 $\pm$ 9.1 \\
    \hline
        2 & 12669 & 3600 & 12548 & 6693 & 102.8 $\pm$ 5.6 \\
    \hline
        2 & 12782 & 3448 & 12648 & 6557 & 105.9 $\pm$ 3.9 \\
    \hline
        2 & 13393 & 4643 & 13244 & 4501 & 34.4 $\pm$ 4.1 \\
    \hline
        2 & 13760 & 6858 & 13620 & 5488 & 95.9 $\pm$ 5.9 \\
    \hline
        2 & 14395 & 9048 & 14201 & 4139 & 36.7 $\pm$ 4.3 \\
    \hline
        2 & 16075 & 7283 & - & - & - \\
    \hline
        2 & 17048 & 6541 & 16811 & 4241 & 75.5 $\pm$ 5.4 \\
    \hhline{|=|=|=|=|=|=|}
        3 & 12043 & 6114 & 11899 & 6044 & 59.7 $\pm$ 32.8 \\
    \hline
        3 & 12456 & 2716 & 12314 & 6938 & 11.4 $\pm$ 1.8 \\
    \hline
        3 & 12996 & 5828 & 12848 & 4600 & 17 $\pm$ 0.8 \\
    \hline
        3 & 13422 & 6517 & 13272 & 5461 & 89.8 $\pm$ 7.5 \\
    \hline
        3 & 13719 & 6800 & 13582 & 6608 & - \\
    \hline
        3 & 14238 & 9184 & 14064 & 5420 & 73.5 $\pm$ 8.4 \\
    \hline
        3 & 15390 & 8680 & 15219 & 6030 & 30.5 $\pm$ 4.2 \\
    \hline
        3 & 15821 & 2386 & 15604 & 4769 & 33.1 $\pm$ 9.9 \\
    \hline
        3 & 16430 & 7518 & 16193 & 5780 & 30.1 $\pm$ 38.1 \\
    \hline
        3 & 17308 & 6521 & 17054 & 5569 & 137.9 $\pm$ 3.4 \\
    \hline
        3 & 18111 & 3542 & 17870 & 4643 & 50.2 $\pm$ 6.9 \\
    \hline
    \end{tabular}
    \label{tab1}
\end{table}
%

%\newpage

\section{Resonator and Coupling Simulation}\label{Sec:Simu}

\begin{figure}[!h]
 \includegraphics[]{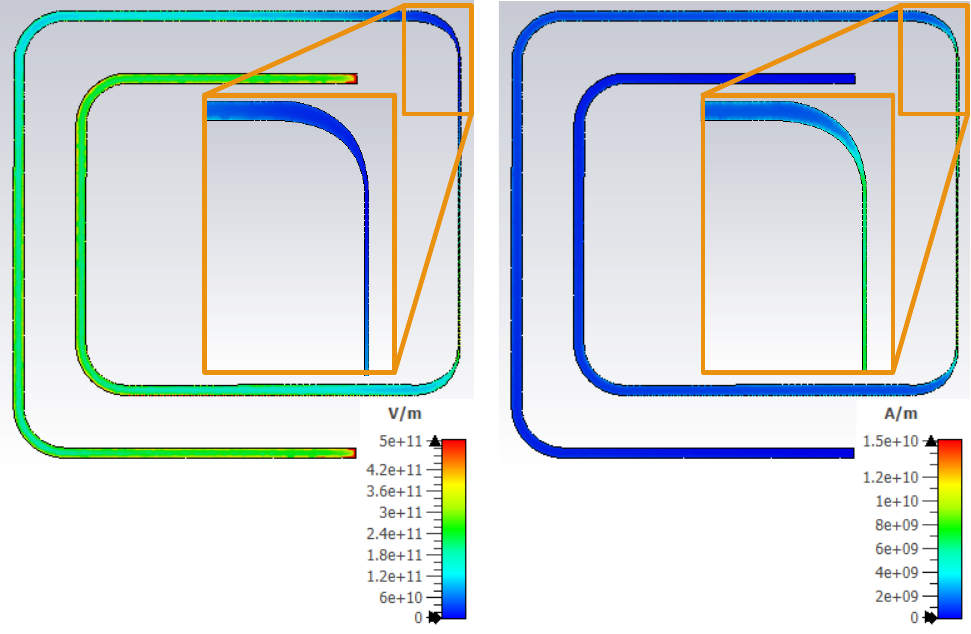}
 \caption{\textbf{Finite element simulations of resonator.} CST Microwave Studio simulation of the distribution of E-fields and H-fields across the resonator structure. The color encoded fields represent the magnitude values.  
 \label{S2}}
\end{figure}

We use finite element and numerical simulations to optimize our resonator design. Key requirements of our resonators are a strong resilience to externally applied static magnetic fields and a small mode volume. To achieve a large field resilience we reduced the area of the resonator to minimize effects of the magnetic field on the superconducting film. Further, we designed the resonators such that they act as lumped element resonators. Here, the resonance frequency is given by the total capacitance and inductance of the structure, with $\wc = \sfrac{1}{\sqrt{LC}}$, analogues to a parallel LC circuit. This allows us to locally separate oscillating electric and magnetic fields and also to concentrate the magnetic fields in more confined regions, resulting in very small mode volumes. To verify the lumped element nature of our resonators we performed finite element simulations, using CST Microwave Studio. Figure\,\ref{S2} shows the resulting magnitude of the E-field (left side) and H-field (right side) distribution along the resonator structure for the resonator design producing the results shown in Fig.\,2 in the main text. The E-field is concentrated along the parallel running wire sections, with its strength approaching zero along the narrow wire section. The opposite is the case for the H-field, where it is zero along the parallel wire sections and strongly concentrated along the narrow wire section. Note, that the H-field magnitude is homogeneous along the whole of the narrow wire section.
%%%
%%%

\begin{figure}[!ht]
 \includegraphics[]{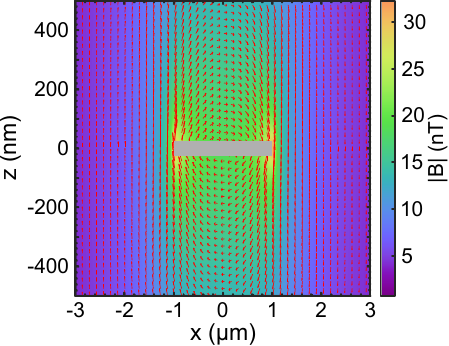}
 \caption{\textbf{Cross-section of resonator magnetic field distribution.} Calculated magnitude of the magnetic field distribution around the cross-section of a rectangular superconducting wire. The wire cross-section lies in the center, indicated by the grey rectangular. The red arrows show the direction of the magnetic field.
 \label{S3}}
\end{figure}

The CST Microwave Studio at hand allowed us a simulation with perfect electric conductors. This is sufficient to model the general electric and magnetic energy distributions and resonance frequencies, however, not to simulate the corresponding oscillating magnetic field distribution, created by a superconducting rectangular wire. To this end, we numerically solve the Biot-Savart law for a rectangular wire cross-section \cite{Primenko2017}, assuming a superconducting current distribution $\textbf{J}_{\mathrm{x,z}}$ \cite{Lee1994},
%%%
\begin{equation}
 \textbf{B}_{1,\mathrm{x,z}} = \frac{\mu_0}{2\pi}\int_{-w/2}^{w/2}\int_{-d/2}^{d/2} \frac{\textbf{J} \times \textbf{r}}{\left(x - x'\right)^2 + \left(z - z'\right)^2} dx'dz',
 \label{eq:B1xz}
\end{equation}
%%%
with the vectors as $\textbf{J} = \left(0,J(x,z),0\right)^T$ and $\textbf{r} = \left(x-x', 0, z-z'\right)^T$ and $\mu_0$ being the magnetic constant. The integration is performed over the cross-section of the wire, of width $w$ and thickness $d$. We define the wire cross-section in the x-z-plane, with $w$ in x-direction and $d$ in z-direction. The length of the wire is along the y-direction. 
For a superconducting wire, the current is not homogeneously distributed over the cross-section of the wire. Current is only flowing on the surface and is exponentially decaying towards the center of the wire. The characteristic length scale is given by the London penetration depth $\lL$. We use the following expression for the current distribution \cite{Lee1994}
%%%
\begin{equation}
J(x,z) = J_1 \left( \frac{\cosh{\sfrac{z'}{\lL}}}{\cosh{\sfrac{d}{\lL}}} \left[ C\frac{\cosh{\sfrac{x'}{l_1}}}{\cosh{\sfrac{w}{l_1}}} + \frac{1 - \sfrac{\cosh{\sfrac{x'}{l_2}}}{\cosh{\sfrac{w}{l_2}}}}{\sqrt{1 - (\sfrac{x'}{w})^2}} \right] + \frac{J_2}{J_1} \frac{\cosh{\sfrac{x'}{\lL}}}{\cosh{\sfrac{w}{\lL}}} \right),
\end{equation}
\label{eq:J}
%%%
where
%%%
\begin{align*}
    \frac{J_2}{J_1} &= \frac{1.008}{\cosh{\sfrac{d}{\lL}}} \sqrt{\frac{\sfrac{w}{\lambda_\bot}}{\sfrac{4*\lambda_\bot}{\lL}} - 0.08301 \sfrac{\lL}{\lambda_\bot}}, \\
    C &= \left( 0.506 \sqrt{\sfrac{w}{2\lambda_\bot}} \right)^{0.75}, \\
    l_1 &= \lL \sqrt{\sfrac{2\lL}{\lambda_\bot}}, \\
    l_2 &= 0.774 \sfrac{\lL^2}{\lambda_\bot} + 0.5152\lambda_\bot, \\
    \lambda_\bot &= \sfrac{\lL}{2d}.
\end{align*}
%%%
The prefactors $J_1$ and $J_2$ define the amplitude of the current density and hence the absolute value of the oscillating magnetic field $B_1$. We define $J_1$ by normalizing the vacuum $B_1$ field to the energy density stored in the resonator \cite{Zollitsch2015, Weichselbaumer2019}
%%%
\begin{equation}
    \frac{1}{2}\frac{\hbar\wc}{2} = \frac{1}{2\mu_0}\int \textbf{B}_1^2 dV = \frac{1}{2\mu_0} B_1^2 V_\mathrm{m},
    \label{eq:B1norm}
\end{equation}
%%%
with $V_\mathrm{m}$ representing the resonator mode volume. The additional factor of $\sfrac{1}{2}$ on the left hand side of \ref{eq:B1norm} takes into account that only half of the total energy is stored in the magnetic field \cite{Schoelkopf2008}. As our resonator design is a quasi 1-dimensional structure we have to define boundaries for the mode volume in the x- and z-direction. A common assumption is to use the width of the conductor wire $w$ \cite{Schuster2007}. For simplicity, we approximate the x-z-area of the mode distribution with the area of an ellipse. For the last dimension we use the length of the narrow wire section, supported by the CST Microwave Studio simulations (see Fig.\,\ref{S2}). In total we find the mode volume to be $V_\mathrm{m} = ( (\pi 3.0\,\micro\meter \times 2.025\,\micro\meter) - w \times d ) \times 300\,\micro\meter = 5696\,\micro\meter^3$.
Figure \ref{S3} shows the resulting distribution of the oscillating magnetic field for the cross-section of the rectangular wire of width $w = 2\,\micro\meter$ and thickness $d = 50\,\nano\meter$. The magnitude $\left| \textbf{B}_{1,\mathrm{x,z}} \right|$ is encoded in the color and the arrows indicate the ${B}_{1,\mathrm{x}}$ and ${B}_{1,\mathrm{z}}$ components of the oscillating field. 
%%%
%%%

With the simulated $\textbf{B}_1$ field distribution we can calculate the position dependent single photon - single spin coupling strength $g_0(\textbf{r})$ \cite{Zollitsch2015, Weichselbaumer2019} for each magnetic moment per unit cell of CGT (ab-plane $0.68\,\nano\meter$ \cite{Li2018, Sun2018}, along the c-axis $0.7\,\nano\meter$ \cite{Gong_Nature2017}). Summation over all CGT unit cells $N$ within the mode volume of the resonator results in the collective coupling strength
%%%
\begin{equation}
    \geff = \sqrt{\sum_{i=1}^\mathrm{N} \left| g_0(\textbf{r}_i) \right|^2} = \frac{g_\mathrm{CGT}\mu_\mathrm{B}}{2\hbar} \sqrt{\sum_{i=1}^\mathrm{N} \left| B_1(\textbf{r}_i) \right|^2} = \frac{g_\mathrm{CGT}\mu_\mathrm{B}}{2\hbar} N_\mathrm{y} \sqrt{ \sum_{i=1}^{\mathrm{N}} \left[ (B_{\mathrm{x,}i}^2 + B_{\mathrm{z,}i}^2) \right] }.
    \label{eq:geff}
\end{equation}
%%%
Here, $\mu_\mathrm{B}$ is the Bohr magneton, $N_\mathrm{y}$ is the number of unit cells along the y-direction and $g_\mathrm{CGT}$ is the g-factor for CGT for which a value of $2.18$ \cite{Khan2019} is used. Note, we give the collective coupling strength for spin $\sfrac{1}{2}$ and for linear polarized microwaves \cite{Zollitsch2015}. 
For the calculation of $\geff$ for the resonator loaded with 15 monolayers of CGT we extracted its lateral dimension from the AFM measurements (see Fig.\,\ref{S3}\,(g)) to $2\,\micro\meter$ along the x-direction and $12\,\micro\meter$ along the y-direction. The flake is assumed to lie directly on top of the superconducting wire without any gap in between. For these values the simulation yields $\geff/2\pi = 8.94\,\mega\hertz$, which is about a factor $2.5$ larger than the experimentally determined value of $3.61\,\mega\hertz$. The overestimation of the simulation most likely results from non-ideal conditions in the experiment. The corresponding flake lies at the top end of the resonators narrow wire section (see Fig.\,\ref{S4}\,(g)), where $\textbf{B}_1$ is concentrated. The finite element simulations show that in this area the field strength is already declining, resulting in a reduced coupling strength. Further, AFM can overestimate the thickness of a flake slightly for when there is a gap between resonator surface and flake \cite{Gong_Nature2017}. The calculation also not includes the multiple peaks observed in the experiment, which - depending on their real nature - can distribute the magnon density over all resonant peaks.
Nevertheless, we can use the simulation to estimate the signal reduction by scaling down the thickness of the flake to a single monolayer. Reducing the simulation to a single monolayer, while keeping the lateral dimensions, results in $\geff/2\pi = 2.33\,\mega\hertz$, a reduction by a factor of $0.26$.

\section{AFM Measurements on CGT Flakes}

%%%
\begin{figure}[!ht]
 \includegraphics[]{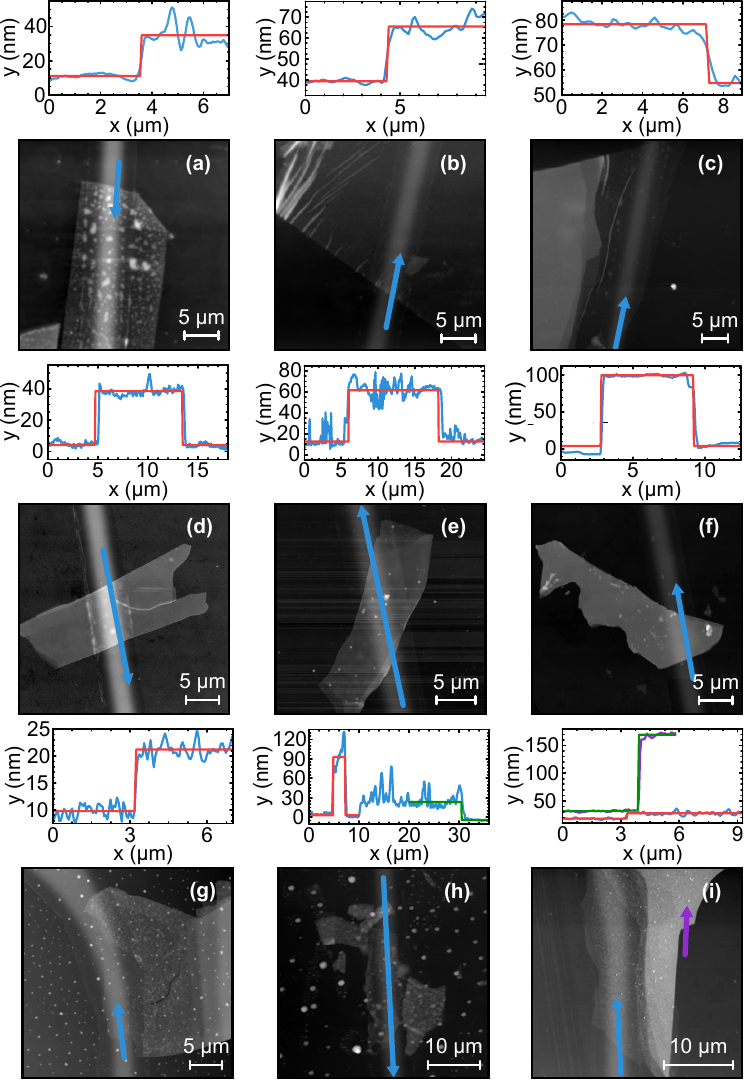}
 \caption{\textbf{AFM measurements.} AFM profile images with respective height profile (above) along the resonator inductor wire (blue and purple lines in profile images, with the arrow indicating scan direction). \textbf{a} - \textbf{c} figures for resonator chip 1 (refer to Tab.\,\ref{tab1}), having resonance frequencies with CGT of $17237\,\mega\hertz$, $17790\,\mega\hertz$ and $16470\,\mega\hertz$, respectively. \textbf{d} - \textbf{f} figures for resonator chip 2 (refer to Tab.\,\ref{tab1}), having resonance frequencies with CGT of $13244\,\mega\hertz$, $12063\,\mega\hertz$ and $13620\,\mega\hertz$, respectively. \textbf{g} - \textbf{i} images for resonator chip 3 (refer to Tab.\,\ref{tab1}), having resonance frequencies with CGT of $12314\,\mega\hertz$, $13272\,\mega\hertz$ and $17054,\mega\hertz$, respectively. The red and green solid lines are fits to the height profiles. 
 \label{S4}}
\end{figure}
%%%
After the transfer of the CGT flakes onto the individual resonators and after measuring FMR, we characterized the thickness of the flakes by AFM. Figure\,\ref{S4} shows a selection of height profile maps from the three resonator chips, including a height profile along the inductor wire of the resonator (blue line in the AFM profile images in Fig.\,\ref{S4}). To extract the thickness we fit the steps in the height profile (red or green lines in the height profiles in Fig.\,\ref{S4}). Note, the height values are relative values with an arbitrary offset. Figure\,\ref{S4}\,(g) shows the thinnest flake of this study, where the processed FMR data is shown in Fig.\,4 in the main text.

\section{Analysis and Additional FMR Data}

\begin{figure}[!ht]
 \includegraphics[]{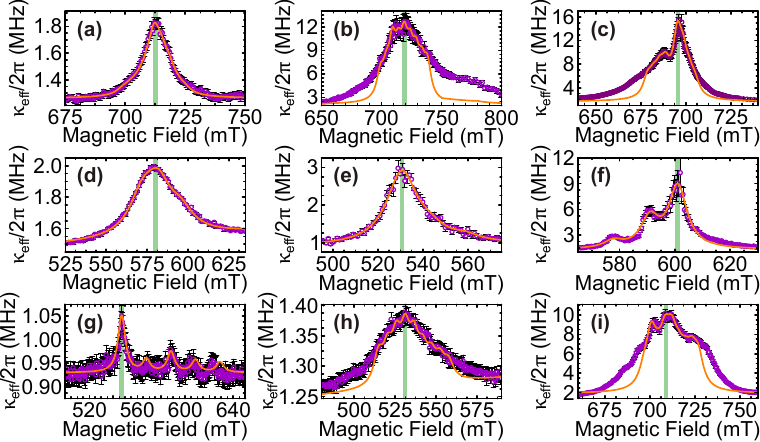}
 \caption{\textbf{Additional data on magnon-photon coupling of CGT-resonator devices.} Results from FMR measurements with effective loss rate $\keff/2\pi$ as a function of the static magnetic field. \textbf{a} - \textbf{c} results for resonator chip 1 (refer to Tab.\,\ref{tab1}), having resonance frequencies with CGT of $17237\,\mega\hertz$, $17790\,\mega\hertz$ and $16470\,\mega\hertz$, respectively. \textbf{d} - \textbf{f} results for resonator chip 2 (refer to Tab.\,\ref{tab1}), having resonance frequencies with CGT of $13244\,\mega\hertz$, $12063\,\mega\hertz$ and $13620\,\mega\hertz$, respectively. \textbf{g} - \textbf{i} results for resonator chip 3 (refer to Tab.\,\ref{tab1}), having resonance frequencies with CGT of $12314\,\mega\hertz$, $13272\,\mega\hertz$ and $11899\,\mega\hertz$, respectively. The orange solid lines are semi-optimized fits, as described in the main text.  The errorbars in the figures represent the standard deviation from the Fano resonance lineshape fit to the respective resonator transmission. 
 \label{S5}}
\end{figure}

We analyze our experimental data, using the model functions (2) and (3) from the main text in a two-step semi-optimized fashion. The main intention for this approach is to minimize the number of free parameters in our model functions. In a first coarse step, we match the collective coupling strength $g_{\mathrm{eff},k}$ to fit the experimental data, assume a constant separation between the individual magnon modes at $B_\mathrm{FMR,k}$ and the same magnon loss rate $\gamma$ for all modes and determine the resonator loss rate $\kappa_0$ from the resonator transmission far detuned from the FMR with the CGT flakes. This results in 3 free parameters for the first stage of our analysis, the magnon loss rate $\gamma$, $B_\mathrm{FMR}$ of the main mode and the constant separation between the $B_\mathrm{FMR,k}$. After this first step we arrive at a best fit to the envelope of the experimental data, however with not matching amplitudes.
In a consecutive second step, we manually optimize the $g_{\mathrm{eff},k}$ to arrive at a model in good agreement with $\wc$ and $\keff$ (see dashed lines in Fig.\,\ref{S5}).

Fig.\,\ref{S5} shows additional results from the corresponding FMR measurements performed on the in Fig.\,\ref{S4} showed resonators. As described in the main text, the measurements were performed at a temperature of $1.8\,\kelvin$ and recording the microwave transmission $\left|S_{21}\right|^2$ as a function of the static magnetic field. Analyzing the microwave transmission by fitting a Fano resonance lineshape to it we extract the effective loss rate of the resonator, interacting with the CGT $\keff$. Figure\,\ref{S5} shows the resulting $\keff$ as a function of the magnetic field. In general, the response of the CGT FMR is complex and varies for the different resonators. The resonance lineshape is not well described by just a single Lorentzian and requires multiple peaks to produce a good agreement. For some resonators, $\keff$ exhibits obvious peaks, residing on a broader spectrum (see Fig.\,\ref{S5}\,(c), (f) and (i)). Together with the observation of well and clearly separated peaks for the resonator loaded with the thinnest CGT flake of $11\,\nano\meter$, we motivating the multiple peak analysis as presented in the main text. However, as the individual peaks are overlapping for the remainder of the resonators we only applied a qualitative analysis. 
%%%
%%%

\begin{figure}[!h]
 \includegraphics[]{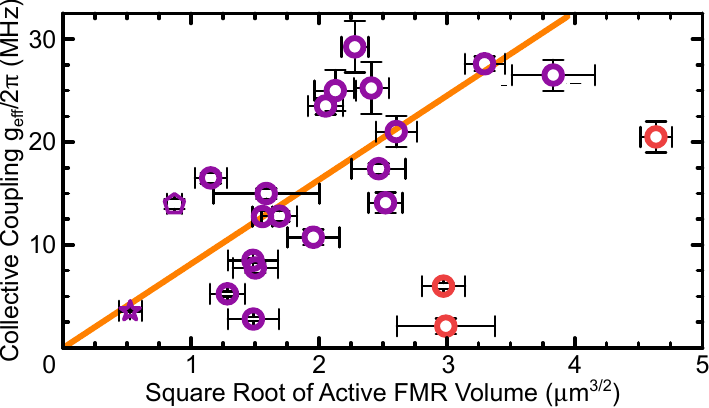}
 \caption{\textbf{Scaling of the collective coupling.} Collective coupling strength $\geff$ as a function of the FMR active CGT-flake volume. The orange line highlights the linear trend of $\geff$ with increasing volume. The red symbols are regarded as outliers, as these flakes show inhomogeneities, leading to inaccurate volume estimations. The star symbol represents data from the thinnest flake (see data in Fig.\,4 in the main text) and the pentagon symbol data from the $17\,\nano\meter$ flake (see data in Fig.\,2 in the main text) The errorbars give confidence values for the extracted values.
 \label{S5.2}}
\end{figure}

Figure \ref{S5.2} shows the extracted collective coupling strength $\geff$ as a function of the square root of the FMR active volume. We define the active volume as the overlap of the oscillating magnetic field $B_1$ and the CGT flake lying on the resonator. The $B_1$ field distribution, discussed in Sec. \ref{Sec:Simu}, is used to estimate the extend of the $B_1$ and is taken as $2\,\micro\meter$. From AFM measurements and microscope images we extract the thickness and lateral dimensions of the flakes to calculate the final active volume. As the collective coupling is proportional to the square root of the number of magnetic moments~\cite{Zollitsch2015}, which are interacting with the resonator field, it follows that $\geff$ scales linearly with the square root of the active volume. This linear trend is highlighted by the orange solid line in Fig.\,\ref{S5.2}. The majority of the extracted data follows this linear trend very well, corroborating our analysis. Only 3 data points deviate strongly from the rest of the data, which we attribute to significant inhomogeneities in the CGT-flakes, making the volume estimation inaccurate. These data points are highlighted in red in Fig.\,\ref{S5.2}.

%%%
%%%

\section{Magneto-Static Spin-Wave Dispersion in Thin-Film Magnets with Perpendicular Anisotropy}

\begin{figure}[!h]
 \includegraphics[]{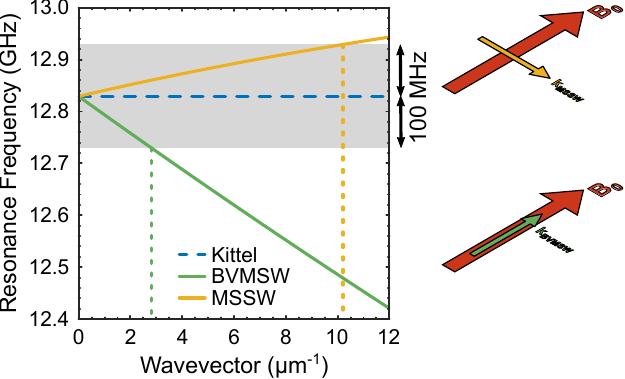}
 \caption{\textbf{Spin-wave dispersion.} Spin-wave resonance frequency for BVMSW (green solid line) and MSSW (yellow solid line) as a function of wavevector. The dashed blue line is the resonance frequency of the $k = 0$ main mode. The parameters used are $B_0 = 598\,\milli\tesla$, $g_{\mathrm{CGT}} = 2.18$, $\mu_0M_{\mathrm{s}} = 194.3\,\milli\tesla$ and $K_{\mathrm{u}} = 3.84 \times 10^4\,\sfrac{\joule}{\meter^3}$ and a thickness of $17\,\nano\meter$. The grey area highlights a $100\,\mega\hertz$ margin relative to the main mode, indicating the order of magnitude of the mode splitting observed in the experiment. The arrows on the right hand side illustrate the relative wavevector orientations of the BVMSW and MSSW spin-wave modes with respect to the static magnetic field.
 \label{S6}}
\end{figure}

Here we describe the spin-wave mode frequency in a thin-film magnet with perpendicular anisotropy along the film normal. We consider this at the magnetic-dipole limit where the wavelength is relatively large and the exchange interaction contribution to the spin-wave dispersion is neglected. Furthermore, standing spin-wave modes along the thickness direction are also ruled out since these modes only appear at much higher frequencies than the main mode, where we consistently observe additional peaks at both higher and lower frequencies from the main mode. The mode (angular) frequency ($\omega$) for wavevector $k = 0$ when we apply a magnetic field $B$ along one of the film plane directions can be given by Eq. 3d in Ref.\cite{Farle1998} as:
%%%
\begin{equation}
    \left( \frac{\omega}{\gamma} \right)^2 = B \left( B + \mu_0 M_\mathrm{s}-\frac{2K_\mathrm{u}}{M_\mathrm{s}} \right).
    \label{kzero}
\end{equation}
%%%
Here, $\gamma$, $M_\mathrm{s}$ and $K_\mathrm{u}$ are the gyromagnetic ratio, saturation magnetization and the perpendicular anisotropy energy density, respectively. Note, that the total field within $\mu_0 M_\mathrm{s}-\frac{2K_\mathrm{u}}{M_\mathrm{s}}$ is negative for perpendicularly-magnetized films which we consider in this section. Within the magnetic-dipole limit, the demagnetization term $\mu_0M_\mathrm{s}$ is modified for spin-waves with finite $k$, depending on the relative orientation between the $M_\mathrm{s}$ and $k$ directions.  Here we follow the expression given in Serga et al. \cite{Serga2010}. For pure backward volume
magnetostatic modes where $k\parallel M_\mathrm{s}$ (illustrated in Fig.\,\ref{S6}), the mode frequency becomes:
%%%
\begin{equation}
    \left( \frac{\omega_\mathrm{BVMSW}}{\gamma} \right)^2 = B \left( B + \mu_0 M_\mathrm{s} \left(\frac{1-e^{-kt}}{kt} \right) - \frac{2K_\mathrm{u}}{M_\mathrm{s}} \right),
\end{equation}
%%%
where $t$ is the thickness of the magnet. Note, that this expression is only valid for the case where $M_\mathrm{s}$ is colinear to $B$, meaning that $\vert B\vert > \vert\mu_0 M_\mathrm{s}-\frac{2K_\mathrm{u}}{M_\mathrm{s}}\vert$. To the limit of $k\rightarrow 0$, the term $(1-e^{-kt})/kt$ is reduced to unity, consistent to Eq.~(\ref{kzero}). When $k$ is nonzero, we can observe that $\omega_\mathrm{BVMSW}$ becomes smaller than that for $k$ = 0, exhibiting a negative group velocity for this spin-wave mode. As the opposite extreme where $k\perp M_\mathrm{s}$ (illustrated in Fig.\,\ref{S6}), the resonance frequency becomes larger than that for $k$ = 0 and is called magneto-static surface spin-wave mode. The mode frequency expression for this mode is given by:
%%%
\begin{equation}
    \left( \frac{\omega_\mathrm{MSSW}}{\gamma} \right)^2 = B \left( B + \mu_0 M_\mathrm{s}-\frac{2K_\mathrm{u}}{M_\mathrm{s}} \right) + \mu_0^2 M_\mathrm{s}^2 \left( 1-e^{-2kt} \right).
\end{equation}
%%%
Here, $\mu_0^2M_\mathrm{s}^2 \left( 1-e^{-2kt} \right)$ is the spin-wave correction term which goes to zero for $k\rightarrow 0$ (hence consistent to Eq.~(\ref{kzero})) and becomes positive for $k > 0$, meaning that $\omega_\mathrm{MSSW}$ becomes larger as soon as spin-waves gain momentum along this direction. We use these two expressions in an effort to explain the origin of the multiple peaks in our experiments.
%%%
%%%
Figure \ref{S6} plots the calculated $\omega_\mathrm{BVMSW}/2\pi$ and $\omega_\mathrm{MSSW}/2\pi$ as a function of wavevector $k$. The range of wavevector is chosen such that the resulting resonance frequencies are within the same order of magnitude as the observed mode splittings in the experiment ($\propto 100\,\mega\hertz$). The corresponding wavelength to a $100\,\mega\hertz$ resonance offset to the main mode are about $2.2\,\micro\meter$ and $620\,\nano\meter$ for $\omega_\mathrm{BVMSW}$ and $\omega_\mathrm{MSSW}$, respectively. These values are within a reasonable scale for our different lateral CGT flake dimensions under investigation. This suggests that spin-wave modes are likely the origin of the multiple resonance peaks observed. 

The thinnest CGT flake shows, however, a deviation from this behaviour. We only observe modes at lower frequencies, which would indicate to BVMSW modes. Calculating the respective shortest wavelength results in $225\,\nano\meter$, which is significantly shorter than for the other devices. We assume that the placement and irregular shape are likely to cause this difference. First, this flake is placed at the very edge of the inductor wire, where the $B_1$ field strength is declining (see Fig.\,\ref{S2}), reducing the FMR active area. Thickness steps can lead to a wavelength down-conversion \cite{Stigloher2018}, however, with the overall irregular shape of the flake it is difficult to define a length scale for a standing spin wave mode.

\section{Atomistic Spin Dynamics Simulations of FMR}

To study the ferromagnetic resonance in CGT we perform atomistic spin dynamics simulations \cite{wahab2021quantum,kartsev2020biquadratic}. The magnetic Hamiltonian employed in the simulations is given by:
%%%
\begin{equation}
\mathcal{H} = -\frac{1}{2}\,  \sum_{i,j }  \mathbf{S}_i \mathcal{J}_{ij} \mathbf{S}_j \:- \sum_{i} D_i (\mathbf{S}_i\, \cdot  \mathbf{e})^2 - \sum_{i} \mu_i  \mathbf{S}_i\, \cdot ( \mathbf{B_0+B_1})
\label{gen_ham}
\end{equation}
%%%
where $i$, $j$ represent the atoms index, $\mathcal{J}_{ij}$ represents the exchange interaction tensor, $D_i$ the uniaxial anisotropy, which for CGT is orientated out of plane ($\mathbf{e}=(0,0,1)$) and $\mathbf{B_0}$ the external static magnetic field applied in-plane during the ferromagnetic resonance simulations and $\mathbf{B_1} = B_1 \sin (2 \pi \nu t)$ the oscillating field applied perpendicular with respect to  $\mathbf{B_0}$. The CGT system has been parameterized from first principle methods\cite{Gong_Nature2017}, up to the third nearest neighbor intralayer and interlayer exchange. The exchange values have also been re-scaled by Gong et al. \cite{Gong_Nature2017} with a 0.72 factor to obtain the experimental $T_C$ and multiplied by $S^2$ to match the magnetic Hamiltonian.  The magnetic moment or Cr is considered 3.26 $\mu_\mathrm{B}$ \cite{Verzhbitskiy2020} and the uniaxial anisotropy has a value of $0.05$ meV as extracted from first principle methods \cite{Gong_Nature2017}. The parameters used in the simulations are given in Table\,\ref{simulation_parameters}. 
%%%
\begin{table}[!h]
\centering
\fontsize{9}{8}
\begin{tabular}{c|l|l|l}
Quantity~ & Symbol~ &quantity~ & units~ \\
 \\\hline
Timestep & $ts$ &0.1 &fs \\
Thermal bath coupling & $\alpha$ &0.02  &\\
Gyromagnetic ratio & $\gamma_e$ &1.760859 $\times$ $10^{11}$  & rad s$^{-1}$T$^{-1}$ \\
Magnetic moment & $\mu_B$ & 3.26\cite{Verzhbitskiy2020} & $\mu_B$ \\
Uniaxial anisotropy            & $D_i$ &0.05 \cite{Gong_Nature2017} &meV/link \\
Simulation temperature            & T &0.001 &K \\
Static magnetic field &$B_{0}$& 0.9, 0.7 & T\\
Oscillating magnetic field amplitude &$B_{0}$& 0.001 & T\\
FMR frequency &$\nu$& varied & GHz\\
Intralayer exchange, NN& $J_1$ & 2.71 \cite{Gong_Nature2017}& meV/link \\
Intralayer exchange, 2NN& $J_2$ &- 0.058 \cite{Gong_Nature2017}& meV/link \\
Intralayer exchange, 3NN& $J_3$ & 0.115 \cite{Gong_Nature2017}& meV/link \\
Interlayer exchange, NN& $J^z_1$ & -0.036 \cite{Gong_Nature2017}& meV/link \\
Interlayer exchange, 2NN& $J^z_2$ &  0.086 \cite{Gong_Nature2017}& meV/link \\
Interlayer exchange, 3NN& $J^z_3$ & 0.27 \cite{Gong_Nature2017}& meV/link \\
\end{tabular}
\caption{Simulation parameters for FMR on CGT system}.
\label{simulation_parameters}
\end{table}
%%%
FMR calculations have previously been employed for atomistic models, and can reproduced well the variation of linewidth with temperature, for example, in recording media systems \cite{Strungaru2020}. Hence, in the current simulations we use the same setup of frequency swept FMR \cite{Strungaru2020}  and we obtain the spectra by performing a Fourier transform of  the magnetisation component parallel to the oscillating field. Since these calculations are done close to 0K, no averaging is require to reduce the thermal noise. To excite the FMR mode, we apply a DC field in-plane of 0.9 T on x-direction and an AC field perpendicular to the DC field, on y-direction. The Fourier transform has been performed for the y-component of magnetisation for $5\,\nano\second$ after an initial $1\,\nano\second$ equilibration time. A thermal bath coupling has been chosen in agreement with the upper limit of the Gilbert damping observed in experiments. 

\begin{figure}[!h]
 \includegraphics[trim={0cm 1cm 10.5cm 0},clip,width=0.75\textwidth]{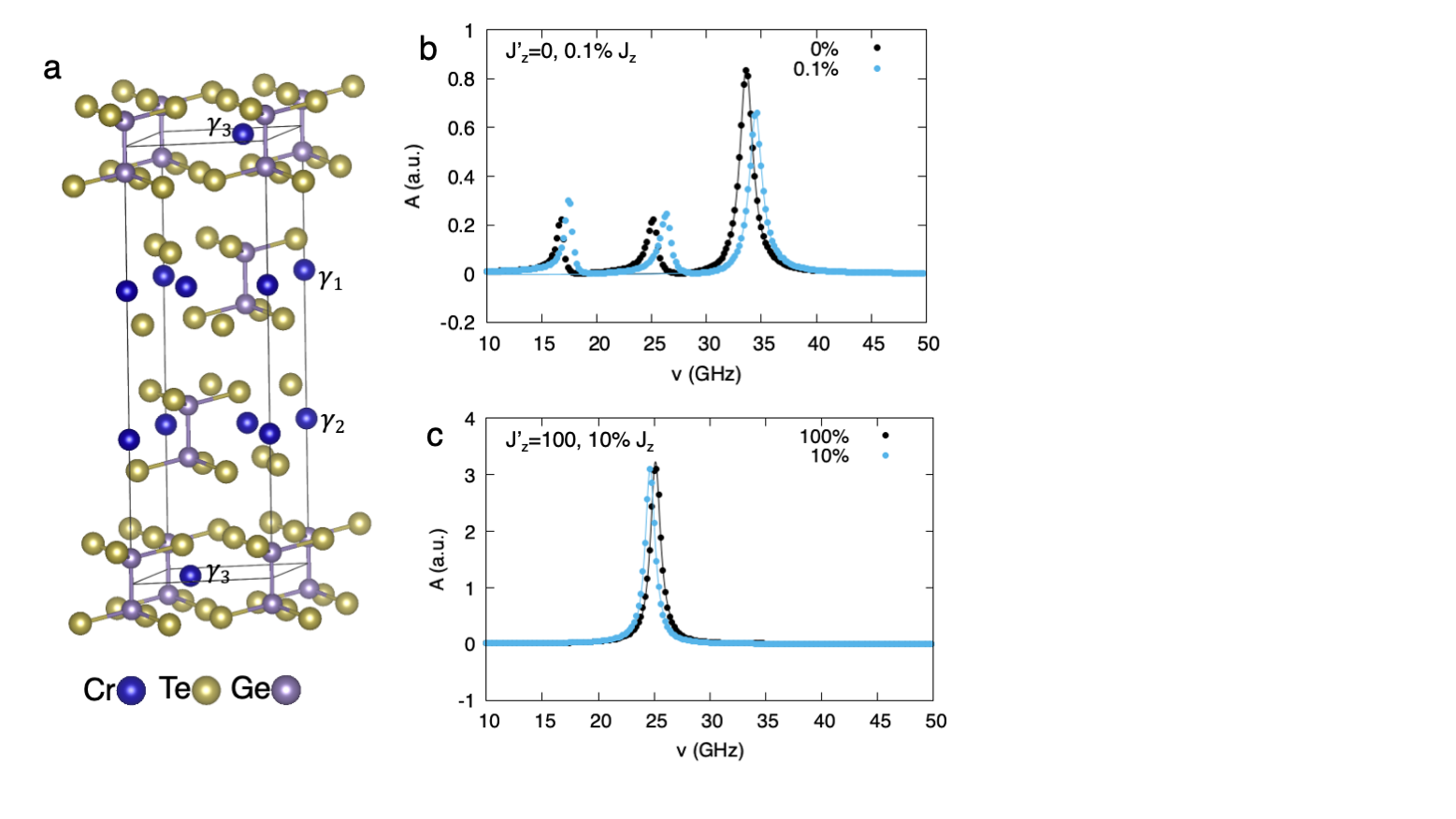}
 \caption{{\textbf{Atomistic simulations.} \bf a,} Schematic of the crystal structure of CGT with atoms defined by different colours. {\bf b,} FMR spectra of 4 layer CGT where the layers are low interayer exchange coupled ($0, 0.1\% ~J_z'$, where $J_z'$ is the pristine CGT interlayer exchange). 
 {\bf c,} Similar as {\bf b,} but with the layers at a strong exchange coupling ($10\%, 100\%~J_z'$). The solid lines in {\bf b-c} represent a Lorenzian fit to the numerical data. \label{fig::cgt_simulations}}
\end{figure}

The system size we performed FMR on is a 4-layer CGT system, with lateral size of $6.91\,\nano\meter \times 11.97\,\nano\meter$, periodic boundary conditions in xy and total of 1600 atoms. The small system size has been used to reduce the computational cost associated with FMR simulations. Experiments have showed modified g-factors due to photon-magnon coupling hence hereby we propose a simple model where the properties of the individual layers have been modified to include different gyromagnetic ratio, as illustrated in Fig.\,\ref{fig::cgt_simulations}{\bf a}. 

We can define the resonance frequencies for each magnetic layer using the Kittel equation in the case of in-plane applied field with perpendicular anisotropy $B_{\perp\mathrm{u}}$:
%%%
\begin{equation}
    \omega=\gamma\sqrt{B_0(B_0 - B_{\perp\mathrm{u}})}
\end{equation}
%%%
We next investigate the FMR signal for a few cases assuming the CGT monolayers at low or strong interlayer exchange couplings  $J_z' = 0, 0.1\%, 10\%, 100\% J_z$, where $J_z$ corresponds to the pristine interlayer exchange (Fig.\,\ref{fig::cgt_simulations}{\bf b-c}). In the low interlayer exchange regime ($J_z' = 0, 0.1\% J_z$), the CGT presents multiple peaks with each frequency corresponding to the layer dependent gyromagnetic ratio, $\gamma$ - $\nu(\gamma_1) = 16.81\,\giga\hertz$, $\nu(\gamma_2) = 25.22\,\giga\hertz$, $\nu(\gamma_3) = 33.62\,\giga\hertz$. 
At $J'_z=0.1\% J_z'$ (Fig.\,\ref{fig::cgt_simulations}{\bf b}) we can still observe resonance peaks corresponding to each individual layer. However by increasing the exchange coupling to $10\% J_z'$ or higher (Fig.\,\ref{fig::cgt_simulations}{\bf c}) there is a single FMR peak indicating that the system  behave coherently with all layers having the same FMR frequency. The single FMR frequency corresponds to the average magnetic properties of the CGT layers. Small variations of the resonance frequency as function of the inter-layer exchange coupling can be observed which these being correlated to the transition of the system from the multi-peaks regime to a coherent excitation. 
By calculating the damping of the highest resonance peaks from a Lorenzian fit, we reobtain the damping corresponding to the input thermal bath coupling, $0.02$ 
with a relative tinny error $\sim5\%$. Overall, the interlayer exchange coupling locks the dynamics of individual layers coherently  together without allowing multiple frequencies at the FMR signal\cite{dataFMR}.  
%%%
%%%

%

\end{document}